# A New Technique of the Virtual Reality Visualization of Complex Volume Images from the Computer Tomography and Magnetic Resonance Imaging

Iva Vasic[♦,✉], Roberto Pierdicca[♦,✉], Emanuele Frontoni[♦,✉], Bata Vasic[♣,✉]

[♦]Università Politecnica delle Marche, Ancona, Italy
vasic@iva.silicon-studio.com, {r.pierdicca, e.frontoni}@
staff.univpm.it
[♣]University of Nis – Faculty of Electronic Engineering, Nis, Serbia
bata.vasic@ppf.edu.rs

**Abstract.** This paper presents a new technique for the virtual reality (VR) visualization of complex volume images obtained from computer tomography (CT) and Magnetic Resonance Imaging (MRI) by combining three-dimensional (3D) mesh processing and software coding within the gaming engine. The method operates on real representations of human organs avoiding any structural approximations of the real physiological shape. In order to obtain realistic representation of the mesh model, geometrical and topological corrections are performed on the mesh surface with preserving real shape and geometric structure. Using mathematical intervention on the 3D model and mesh triangulation the second part of our algorithm ensures an automatic construction of new two-dimensional (2D) shapes that represent vector slices along any user chosen direction. The final result of our algorithm is developed software application that allows to user complete visual experience and perceptual exploration of real human organs through spatial manipulation of their 3D models. Thus our proposed method achieves a threefold effect: i) high definition VR representation of real models of human organs, ii) the real time generated slices of such a model along any directions, and iii) almost unlimited amount of training data for machine learning that is very useful in process of diagnosis. In addition, our developed application also offers significant benefits to educational process by ensuring interactive features and quality perceptual user experience.

# 1 Introduction

The past decade has witnessed a steady increase in multi-dimensional representations of medical models and processes utilizing advanced concepts in Artificial Intelligence (AI) to enhance capabilities of the existing VR and AR software tools [1, 2] in variety of applications such as diagnosis, pre-operation planning, various simulations, and also in educational and clinical trainings [3, 4, 5].

Increasing the efficiency of in-vivo diagnostics and consequently on adequate educational techniques relies on equipment/hardware that uses the knowledge of the anatomy of human bodies giving the both semantic and visual information [6]. As a representative example CT and MRI systems with developed image processing software are continuously refined to provide very exact visual information of even tiny parts of the human body [7]. However, due to technological limitations even the superiority of above mentioned expensive hardware is not sufficient to provide completely accurate information. Imperfections are mostly related to relatively small resolution of scanned images as well as in the insufficient number of scanning steps (number of the obtained image slices). This problem is solved by employing software algorithms based on image processing that interpolate intermediate steps and uneven attenuation of different tissue types [8]. The State-of-the-art hardware equipment and software improvements still do not meet all the requirements of medical doctors with even extensive experience in traditional diagnostic. This is exactly the where new VR and AR techniques provide significant benefits.

In response to new requirements, the automatic 3D model generation has been implemented in standard CT and MRI scanner software packages based on mathematical Radon transformations [9]. Although these improvements have shown measurable diagnostic results, the automatically generated 3D models have contained significant geometric and topological irregularities to be completely usable in its source form. In order to meet the visual criteria for displaying 3D models, lot of visualization techniques have been proposed. However, such methods are mainly based on Computer Aided Design (CAD) and manual construction models and computer animations using existing software [10]. This approach enables impeccably high-quality representation of the model, where the model usually does not correspond to the real geometric details of the CT and MRI models.

Our approach combines 3D mesh processing process and software coding within the gaming VR engine using complex volumetric images obtained from CT and MRI. Algorithm operates on real representations of human organs without any structural approximations of the real physiological shape. In order to obtain realistic representation of the mesh model geometrical and topological corrections are performed in the term of smoothing and materialization of the mesh surface with preserving real shape and geometric structure. Using simple mathematical intervention on the 3D model and mesh triangulation the second part of our algorithm ensures an automatic construction of new 2D shapes that represent vector slices along any user chosen direction. The final result of our algorithm is developed software application that allows to user complete visual experience and perceptual exploration of real human organs through spatial manipulation of their 3D models Thus our proposed method achieved threefold effect: i) high definition VR representation of real models of human organs, ii) the real time generated slices of such a model along any directions, and iii) practically unlimited source of training data for the AI learning. In addition, our developed application also offers significant benefits to educational process by ensuring interactive features and quality perceptual user experience.

The paper is organized as follows. The Section 2 describes the general notations and defines main group of problems in the VR visualization of real models of human organs as a consequence of errors in segmentation of the volumetric images from CT and MRI devices. In this section we also briefly describe achievements of several known methods that give solutions in overcoming particular visualization issues. Our algorithm is described in detail within the Section 3. This section also contains theoretical background of used technique with mathematical and geometrical notations. All steps of proposed algorithm are explored with developed software codes used for 3D mesh model manipulation and transformation. Experimental results with visual illustration of our method contribution are shown in the Section 4. The brief conclusion and the further work directions are given in Section 5.

## 2 General Notations and Problem Definition

All modalities of medical imaging techniques are widely used and very effective in diagnosis, treatment planning, and evaluation especially in qualitative and semi-quantitative inspections performed by experienced medical doctors. However, the complex features of particular medical images can complicate the analysis and also slow down a successful diagnosis. In order to make the complex characteristics of the medical picture acceptable and perceptually understandable, defining anatomy using 3D geometric modeling and visualization becomes necessary in clinical practice, but also in the fields of education.

Modern technologies in the field of medical imaging have reached a multitude of general requirements in diagnosis until the emergence of new visual criteria in almost all areas of clinical model analysis [11]. In order to recognize the overall problem of the concept of visualization 3D geometric structures obtained from CT and MRI scanners we defined the schematic structure of the visualization process (**Fig. 1**).

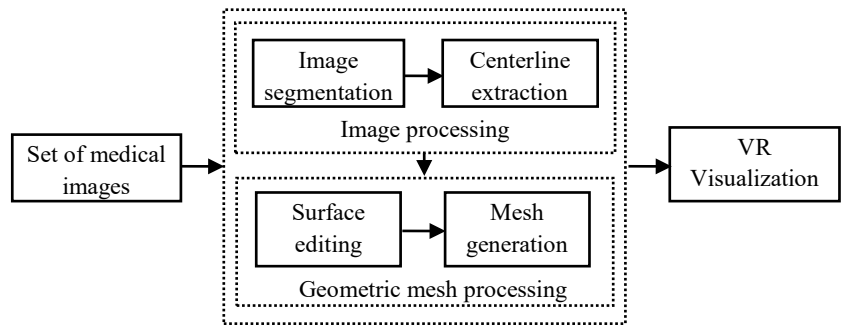

**Fig. 1.** VR visualization process flow

The abundant number of images obtained from CT or MRI scanners ensures the visualization by segmentation particular tissue types and modeling of the 3D geometric structures. Regardless to usability of scanned images itself, in this paper we concentrate to prove that generated 3D structures with some geometric interventions are fully useful for visualization and educational concepts, and also spur ideas for scientific or clinical predictions.

### 2.1 Image Processing Problems

The first group of problems is related to CT and MRI hardware and quality of processing software tools. All images from the set are distinguished from each other by the attenuation value of scanned tissue (X-ray absorption and scattering in CT scanning and static X-ray imaging), dimension, type of motion, and timescale and periodicity of the motion (in the case of CT and MRI scans of the non-static anatomic structures). The balance of signal-to-noise ratio and spatial resolution affect the quality of the image and consequently the amount of image editing and segmentation before the visualization. Good visualization depends also on the choice and affordability of appropriate image modality, but this group of issues is actually the radiologists' field of research and we will not dig deeply into this area. Since our point of interest is the visualization, we will consider that the whole image processing and segmentation part of processes are performed satisfactory precise.

At the beginning of the whole process shown in **Fig. 1** is the set of slice images along the main axes in 3D space. The standard CT and MRI format of the scanned image set is DICOM (Digital Imaging and Communications in Medicine). The example of arbitrary selected slice of the CT scanned heart structure is shown on **Fig. 2** using ITK-SNAP software [12].

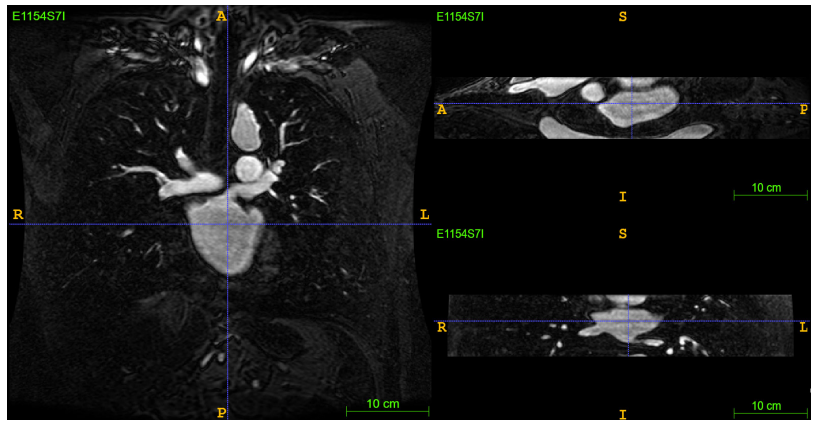

**Fig. 2.** DICOM slice images of the heart across coronal, sagittal and transverse planes

Segmentation and centerline extraction within the image processing step are covered by the common standalone software engines for the medical 2D images visualization: ITK-SNAP [12], MITK [13], and 3D Slicer [14]. However, the quality of the obtained 2D volume images is often insufficient for the construction of satisfactory precise 3D structures, which are actually a demand of challenging medical diagnosis.

### 2.2 Mesh Generation Problems

Previously mentioned problems require more profound involvement of notable mathematical functions such as Radon transform for the generation of 2D vector shapes and 3D structures. The Radon transformation $R_f$ is a function defined on the space of straight lines $L \in \mathbb{R}^2$ by the line integral along each line [10]:

$$R_f(L) = \int_L f(x)|dx| \qquad (1)$$

The function $f(\mathrm{x}) = f(x, y)$ is a continuous function, which the double integral $\iint \left( |f(\mathrm{x})| / \sqrt{x^2 + y^2} \right) dxdy$ extending over the whole plane converges, and it holds that:

$$\lim_{r \to \infty} \int_0^{2\pi} f(x + r\cos\phi, y + r\sin\phi)d\phi = 0 \qquad (2)$$

Visualization plugins of CT and MRI supporting software packages already relays on some similar functions. However, post-processing techniques are unable to completely solve the issue of irregular meshes that CT and MRI provide. The next **Fig. 3** illustrates the real 3D STL (stereo-lithography) model of the heart, and detail of its mesh area with topological and geometrical errors.

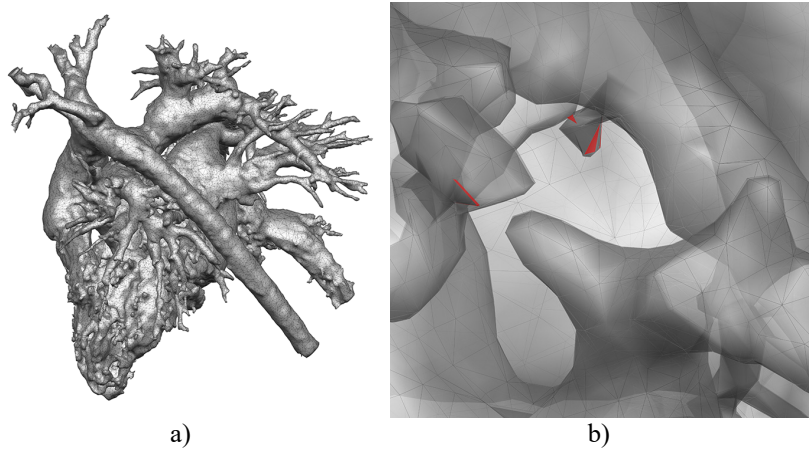

**Fig. 3.** 3D heart model: a) entire mesh, b) topological errors

Such errors have strong influence on the further operations of visualization in general, and particularly on the materialization process. This negative influence decreases an overall perceptual visibility of the structures of interest. In order to find an appro-priate solution in solving structural geometrical issues, we can define two crucial types of topological errors: i) isolated vertices and structures, and ii) shared geometric primitives (vertices and faces) that belong to multiple anatomic structures. Both types of mesh errors we may consider as a result of imperfection of image processing algorithms and methods. In other hand we have to observe such results from the medical point of view. In this context, some of isolated and/or shared structures may be part of real anatomic structure "invisible" to artificial eyes that hardware equipment represent.

In contrary to problems described in previous subsection where image processing algorithms tries to improve accuracy and increase scanning resolution, according, according to our knowledge there is no automatic methods that can successfully solve geometrical problems of already generated meshes. However, some of the existing approach can be very useful in the process of refining the mesh geometry and removing topological errors [15]. Corrections of the second type of errors require medical consultation and semi-automatic and even manual intervention on the mesh geometry. The combination of existing techniques and the improvement of appropriate algorithms can lead to the successful overcoming of problems and geometric structures that meet the criteria of visualization.

### 2.3 Training Set for Machine Learning

Achievements of AI achievements are mostly inspired from the nontrivial problems in medical sector and human biology. The strength of powerful neural networks (NN) always relies on rich training data. Although some of the medical areas have abilities to ensure lot of information, the lack of information in other medicine fields is evident. Insufficient data availability can be often a reflection of a complex and expensive technological process of diagnosis, but also of inefficient, unsafe, and uncertain invasive diagnostic techniques.

CT and MRI imaging technologies are classified as very expensive diagnostic methods, so the amount of available scanned data is very limited, especially when the order of magnitude of the data required for machine learning is taken into account. The power of AI was mainly used in the processes of image generation and segmentation [16], while the results were insufficiently used in the 3D geometry analysis and education. In these segments our approach offers possibilities in collecting almost unlimited amount of data for the machine learning and thus significant improvement of the both, diagnostic and education fields.

## 3   Our Algorithm

Our method is focused on the three research targets: i) high definition VR representation of CT and MRI generated volumetric models, ii) the slices generation along any direction, and iii) obtaining the training data for the machine learning in medical diagnosis purpose. According to these emphases, our algorithm proposes the complete framework of mesh processing and visualization steps presented in **Fig. 1**.

All including processes with relevant operations and techniques are presented in the flowchart (**Fig. 4**) with introduced general notations of all used variables, features and functions.

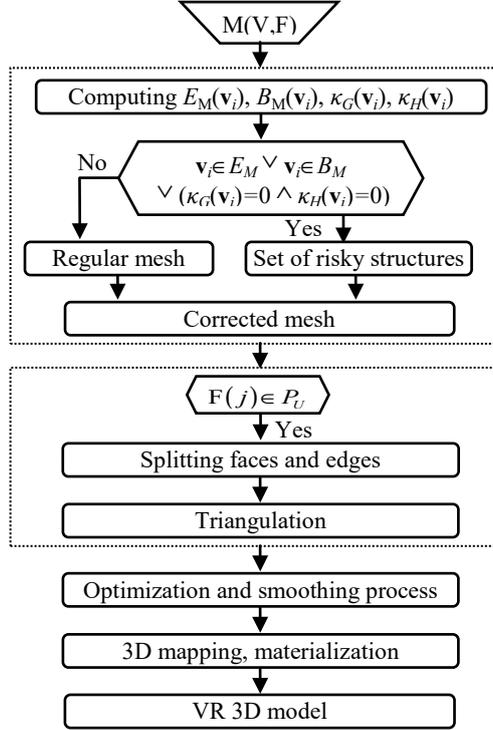

**Fig. 4.** The flow chart of our algorithm

The input of proposed method is the CT or MRI generated mesh of a given 3D surface $M(V,F) \in \mathbb{R}^3$, where $V(v_x, v_y, v_z)$ and $F(i)$ are respectively matrices of all Cartesian vertex coordinates and faces constructed by indexes $i = 1, \cdots, n$ of $n$ belonging vertexes.

### 3.1 Mesh Correction

In order to locate mesh errors and undefined isolated structures, the first step includes the main computations of geometrical and topological features. Within this step algorithm extracts the matrix of the topological error vertices $E_M$, and the matrix of boundary vertices $B_M$ [1] for each vertex $v_i(v_{xi}, v_{yi}, v_{zi})$. Using these matrices as well as the sign and a threshold value[2] of $\kappa_G(\mathbf{v}_i)$ and $\kappa_H(\mathbf{v}_i)$ we first extract all risky vertices and structures. The matrix of risky vertices is actually an union of matrices EM, BM and a set of vertices which satisfy the condition: $\kappa_G(\mathbf{v}_i) = 0 \wedge \kappa_H(\mathbf{v}_i) = 0$. The result of this step is the set of isolated geometric structures of irregular topological properties and shapes, but the recognition and classification of specific anatomical structures are excluded from an automatic processing. Their importance for the further process is estimated by medical doctors.

---

[1] Boundary vertices are often important for shape creation, and algorithm leaves to user a choice of their removing from the mesh.

### 3.2 Slicing the Mesh Along User Direction

This part of algorithm is strongly related to the geometric operations and computations over the mesh surface and their primitives. Quality of CT or MRI generated geometric structure is crucial for all further computations and results. The internal structures of scanned organs are especially important because the accuracy of our representation depends on their definition. In order to theoretically describe the details of our idea, we assume that the scanned model contains all internal structures and that they are topologically regular.

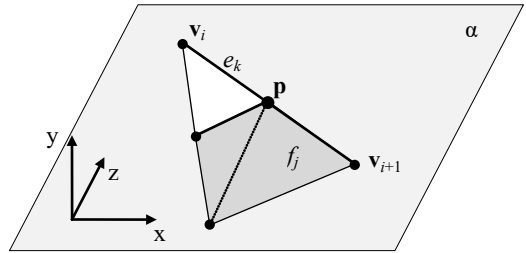

**Fig. 5.** Mesh face $f_j$ intersected by the plane $\alpha$

For the given regular mesh $M \in \mathbb{R}^3$ we define set of triangular faces $F(f_1, \cdots, f_m)$ constructed by the three vertex indices associated to their Euclidean coordinates. Consequently, each triangle face $f_j$ of the mesh surface is defined by the coordinates of corresponding vertices. Each of the two corresponding vertices virtually forms an edge between them (**Fig. 5**). The intersection point $\mathbf{p}(x,y,z)$ of the edge of interest $e_k(\mathbf{v}_i, \mathbf{v}_{i+1})$ and the desired plane $\alpha : ax + by + cz = 0$ is calculated by solving the pair of simple plane and edge equations. In our case the intersection plane is fixed along an axis defined by the user, whereas the whole mesh is aligned and rotated according to this axis. Intersection plane thus have a simple form: $\alpha' : y = s$, where $s$ is a numeric value acquired in real time from the user mouse roll position. Coordinates of intersection point of the plane $\alpha'$ and edge between vertices $\mathbf{v}_i(v_i^x, v_i^y, v_i^z)$ and $\mathbf{v}_{i+1}(v_{i+1}^x, v_{i+1}^y, v_{i+1}^z)$ are calculated as follows:

$$\frac{x - v_i^x}{v_{i+1}^x - v_i^x} = \frac{y - v_i^y}{v_{i+1}^y - v_i^y} = \frac{z - v_i^z}{v_{i+1}^z - v_i^z} = s \qquad (3)$$

---

[2] Threshold values of all elimination criteria, including Gaussian and mean curvature values, are adjustable.

Each face of the mesh which correspondent edges intersect with given plane needs to be divided to new three triangular faces[3] according to intersection points. The new edge between two intersection points defines the segment of the whole intersection spline, which is actually defined segment of our new slice.

If we take into account that the user can arbitrarily rotate the mesh model and manage the positions of the intersection planes in 3D space, the number of cross sections of the model is practically unlimited. Collecting all of resulting slices with the corresponding diagnosis data we can form huge training set for machine learning.

### 3.3 Optimizing Mesh for Visualization

Although CT and MRI techniques produce volumetric images that can be used in monochromatic 2D and 3D visualization, in the VR applications, such a solid models are not perceptually attractive and cannot be used in their original form [17]. In addition, geometry and topology are often defined by sharp faces and shapes, which do not correspond to real shapes of organs. Within the final step our algorithm applies optimization [18] and simplification operations to prepare complex geometric structures for real-time manipulation. On the other hand, we use a semi-automatic mapping and materialization method to achieve a smooth and perceptually acceptable final result of the mesh model.

Due to the insufficient invariance of the geometric structures to some of the methods in this part of the algorithm, each process is performed with constant control and evaluation by specialized medical doctors and with strict respect for preserving natural forms of anatomy and functionality of visual simulations.

### 3.4 VR for Education

In addition to a significant contribution to diagnostics, the proposed method introduces innovations in the educational approach. The perceptually refined model is imported into a developed gaming VR application, which raised the visual experience to a new level. This approach provides measurable knowledge acquisition through interaction between students and our smart VR application, on the one hand, and enrichment of application content using artificial intelligence on the other.

---

[3] The face is not divided if the one of vertex belongs to the intersection plane.

## 4      Experimental Results

In the experimental phase of our work, we used geometric structures, obtained from some of well-known segmentation software, which we mentioned in the Section 2.1. The result of such semi-automatic process is very commonly a generated 3D mesh in the STL file format that is actually native to the stereo-lithography 3D printing technique. As an experimental geometrical structure, the real heart model is loaded in Matlab software where we performed mesh inspection using our developed software functions [15].

Calculating the main features noted in **Fig. 4**, all risky primitives are located and selected over the whole mesh, which finalized the process of forming both matrices: matrix of the topological error vertices $E_M$, and the matrix of boundary vertices $B_M$. Geometrical and topological errors in forms of isolated vertices, edges and faces are immediately removed from the mesh, whereas other risky types of geometrical structures, such as elongated faces, boundary edges and isolated sub-structures, are located and corrected/removed according to the heart anatomy literature [19] in strict consultation with skilled medical doctors - cardiologists or/and cardio surgeons.

### 4.1      Geometrical and Topological Modification of 3D Mesh Model

Mesh simplification and optimization procedure is performed in next task in order to avoid post-scanning computations within different applications and tools that we use for mapping and materializing processes. The main criterion in choosing a level of simplification is preserving the shape of the mesh model that is actually a key demand of the visualization and diagnostic globally. In this task whole mesh area paved by rough triangle faces is additionally relaxed with the mesh smoothing algorithm. In order to meet the perceptual requirement of visualization and achieve satisfactory user experience, realistic textures are produced and assigned to the virtual model. Left (a) and right (b) images in the following **Fig. 6** show the 3D model of the heart before and after refinement respectively.

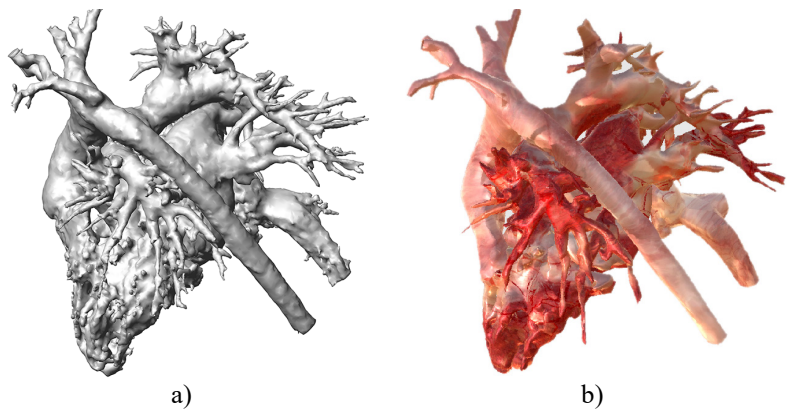

a)                                                                 b)

**Fig. 6.** 3D heart model: a) source model from CT scanner, b) refined and texturized VR model

Unlike the mixed photogrammetry and video technologies, which is a conceptual base of the majority of conventional 3D scanners, the nature of CT and MRI imaging technologies cannot ensure photographic recordings of scanned surfaces. Moreover, these technologies are not constructed to provide any colour information of the scanned tissues. This fact indicates all technologies: optimization; correction; refinement; smoothing; mapping; and materialization as mandatory tasks within any algorithm for visualization of CT and MRI scanned models.

### 4.2 3D Visualization Using Gaming Software Engine

Using all previous tasks our algorithm produced 3D mesh model that is suitable for the high quality visualization. The next action in proposed method goes toward to ensure the interactive features that are useable for both educational purposes: teaching and learning. In order to provide the quality user experience these features in our application [20] are written in C# programming language and developed within Unity engine [21].

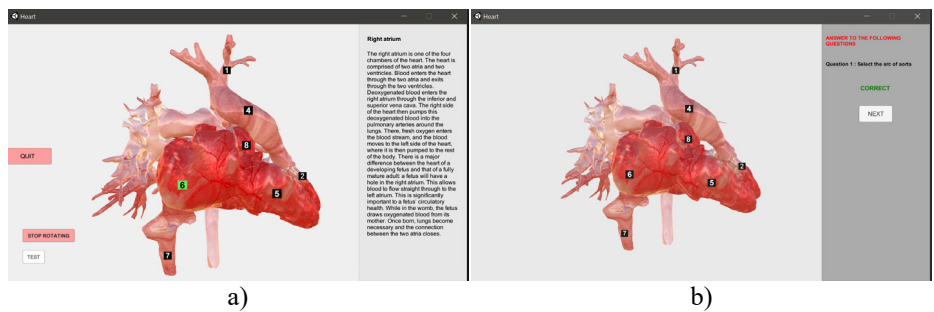

a)                                      b)

**Fig. 7.** Application preview: a) display when one of the annotation numbers is selected, b) display of the quiz mode

Although the all gaming engines have been mostly used for interactive design and development in the gaming industry, we proved in this paper their successful employment in scientific medical and education fields. We have communicated all their results and ideas visually. Our developed application offers of the following interaction possibilities:

- Observation of all perceptual details of any inserted medical model by using the computer mouse functions. All required user actions are simplified in order to provide easy way to perform manipulation without any previous experience. The standard commands are developed: the left mouse key is used for object rotation, the middle one for moving the camera along screen axis, and the scrolling action executed zooming in or out action.
- Obtaining information about the desired part of the medical model. Each part is related to the particular annotation field represented by small quadratic shape and a unique number. By clicking any of these very noticeable numbers on the right hand side of the screen, the textbox with the corresponding anatomic description appears (**Fig. 7** a).

- Testing and evaluating knowledge of students from all previous observations using the quiz mode. This application mode contains interactive fields with questions and several offered answers per each question. Testing software code compares selected choice with the right answer and shows the test result in the form of visual information (**Fig. 7** b).

### 4.3 Comparative Results of Mesh Slicing Methods

In most of cases diagnostic process can be successfully performed using the standard method for observing CT and MRI volumetric models by scrolling in depth along transverse, sagittal and vertical axes [12][14]. However, in remarking anomalies and analysing certain parts of cardiac anatomy such as aorta, more convenient viewing angle is required. Our method provides slicing along any direction, and thus meets the requirements for the shape assessment and measurements.

In order to demonstrate measurable comparison between visualizing methods, 3D heart model is used as an example. Imperfections or distortions of certain slice views are easily observed on relatively regular shapes, so we separated the aortic area from the rest of the model to emphasize all differences. The correct shape of the aortic section and the advantage of our approach in relation to standard software diagnostic tools are clearly perceptible in the following figure **Fig. 8**.

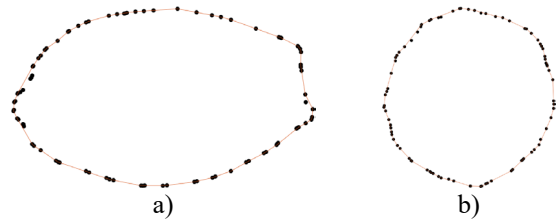

a)        b)

**Fig. 8.** 2D shapes of the aorta 3D model obtained from: a) the standard transverse slicing plane; b) Our method - $30^0$ rotated transverse slicing plane.

The elliptic shape (**Fig. 8** a) of the model intersection corresponds to the output of the existing slicing methods [12][14], whereas our algorithm provides the closely circular section (**Fig. 8** b) of the aorta. Therefore, it is obvious that our approach ensures greater accuracy of assessment and analysis, and thus consequently correct diagnosis. Additionally to the picturesque evidence of the accuracy and suitability of aortic dissection obtained from our algorithm, **Table 1** gives a comparison of numeric values of the aortic features that are important in remarking anomalies and diseases.

Table 1. The aorta mean diameter values (in mm)

| Case 1 | Case 2 | Standard tools | Our method |
|--------|--------|----------------|------------|
| 24.8   | 24.3   | 26.75          | 24.5       |

The first column value (Case 1) represents mean diameter of the normal ascending aorta commonly used in the medical literature [22]. The value in second column (Case 2) is the mean value of the MRI examinations of the normal thoracic aorta of 66 subjects aged in range 19.1–82.4 years [23]. The result for each person is obtained from four measurement made in the axial plane (supravalvular, at the mid-part of the ascending aorta, at the level of the arch on the ascending and descending aorta), and one in a sagittal plane (at the top of the arch).

Our approach superiorly provides calculation of all important features, including diameter, along the entire aorta with just one measurement. The value in the fourth column we obtained by calibrating the slicing plane in order to be always perpendicular to the aortic flow axis. Assuming the 3D model is calibrated to measure the mean aortic diameter in the value range given from the literature, we clearly notice a considerable deviation of the values in one measurement of standard slice tools (Column N°3).

### 4.4 Experimental Test of Our Algorithm

Our algorithm is experimentally tested on various arbitrary 3D models[4] with completely different geometrical and topological structures, shapes and complexity.

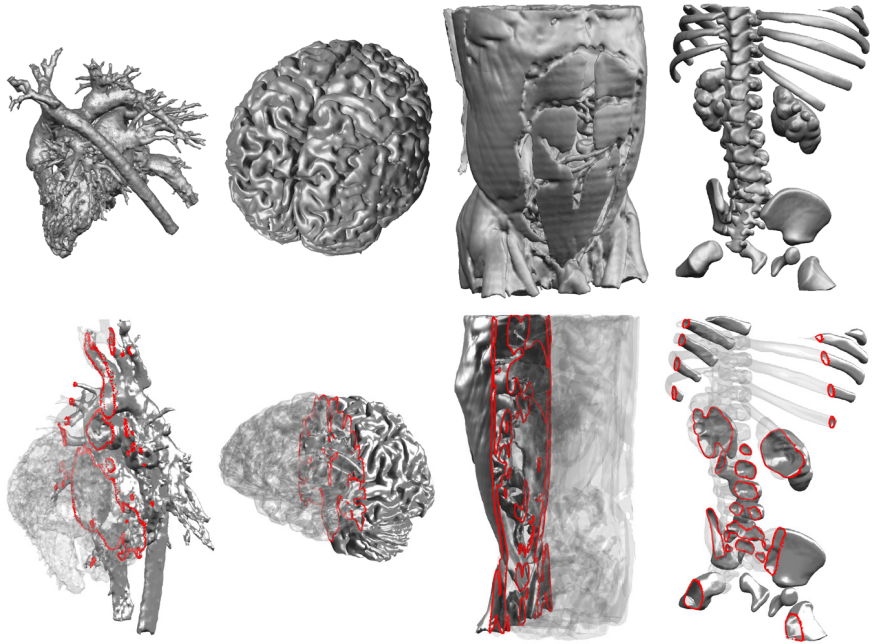

**Fig. 9.** Our algorithm in action: renderings and slicing results (red lines)

---

[4] 3D models are downloaded from the website Embodi3D [24]

The visual results and precision of the method are shown in the previous figure (**Fig. 9**), which also illustrates equally successful operation on compact meshes and complex 3D geometric structures. In addition, we give in **Table 2** numerical features of original and optimized 3D models; and numerical values obtained before and after slicing, as well as slicing speed for each model. It can be noticed that our algorithm is pretty fast and operates on a complex model of kidney and spine in 14 seconds.

**Table 2.** Comparative results of the mesh slicing procces over different 3D models: DC – Coordinate of a slicing plane, T– The slicing computation time

| 3D model | Complexity | DC (x, y, z) | # Vertices | #Faces | T (sec) |
| --- | --- | --- | --- | --- | --- |
| Heart | Original | 0, 80, 0 | 122,408 | 245,962 | 4 |
|  | 50% Optimized |  | 61,204 | 123,530 | 2 |
| Brain | Original | 0, 50, 0 | 332,152 | 664,988 | 7 |
|  | 50% Optimized |  | 166,076 | 334,120 | 4 |
| Torso | Original | 0, 30, 0 | 497,751 | 996,198 | 10 |
|  | 50% Optimized |  | 248,875 | 498,446 | 6 |
| Kidney and spine | Original | 0, -220, 0 | 534,394 | 1,068,777 | 14 |
|  | 50% Optimized |  | 267,197 | 534,383 | 7 |

### 4.5 Theoretical Model of Machine Learning Using 3D Model Slicing

This paper introduced a new idea of use a virtual application in the AI learning process and presented the new theoretic method of the training set forming. Although the construction of NN has significant influence on its efficiency, in this paper we focused on the training data quantity and quality. In addition to numerous effects of our VR application to medical diagnosis and education, it has also strong potential in the machine learning, particularly in training set forming.

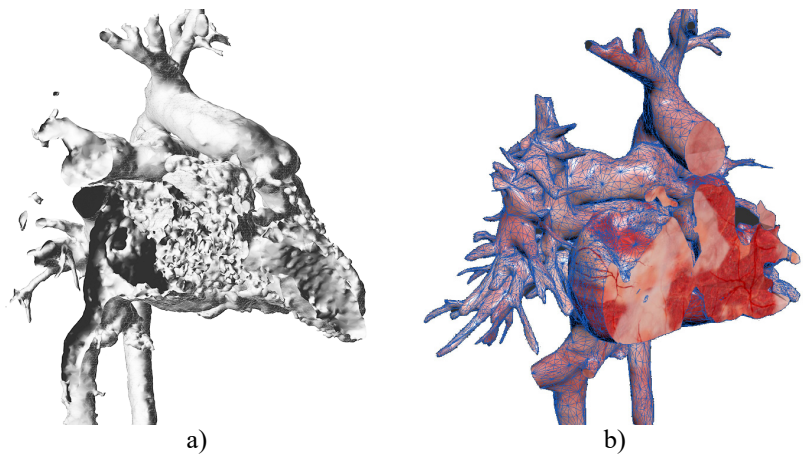

a)            b)

**Fig. 10.** Slices of the 3D model along Y axis: a) Matlab isosurface, b) Unity application slicer

According to simple mathematical equations given in Section 3.2 we preliminary developed an example procedures for 3D mesh slicing using Matlab isosurface and C# functions. The result of both techniques respectively is shown in the previous figure (**Fig. 10**).

Both slice images are associated to the same 3D model and to same medical case or diagnose. Thus, all different camera views in combination with slices along all direction are potential source of almost unlimited number of training data.

## 5    Conclusion and Future Works

In this paper, we presented a novel technique that offers a systematic method for the VR visualization complex volume images from the CT and MRI imaging devices. We combined 3D mesh processing and software coding within the gaming engine. Experimental practical results showed real VR representation of human organs without any structural approximations of the real physiological shape. The second part of algorithm ensured an automatic construction of new 2D shapes using mathematical intervention on the 3D model and mesh triangulation. Constructed shapes represent vector slices along any user chosen direction. As the final result we developed software application that allows to user complete visual experience and perceptual exploration of real human organs through spatial manipulation of their 3D models.

We also introduced the new theoretical method of use a virtual application in the machine learning process. Experimental simulation proved significant potential of our algorithm and almost unlimited amount of training data for machine learning that is very useful in process of diagnosis and education. The limitations of the proposed technology are only reflected in the small number of processed CT and MRI models. However, our future works will include using a bigger set of different 3D models and improving overall performance gain, by developing the additional software procedures for real time production of slice images for NN training. Our further research will also strive to cover more medical areas and provide relevant information, and also expand the area of interest to other educational and research fields.

**Acknowledgments** The authors sincerely acknowledge the information and suggestions provided by the cardiologist Miomir Randjelovic, Cardiovascular diseases clinic in Nis, Serbia.